\documentclass[twocolumn]{IEEEtran}
\usepackage{latexsym}\usepackage{graphicx}
\newcommand{\psfx}[2]{\includegraphics[width=#2]{#1}}
\newcommand{\bfx}{\mbox{\boldmath $x$}}

\newcommand{\beq}{\begin{equation}}\newcommand{\eeq}{\end{equation}}

\newtheorem{exam}{{\bf Example}}
\usepackage{xfloat}\usepackage{cite} \usepackage{amssymb}\usepackage{amsmath} 
\begin{document}
\clearpage\pagestyle{empty}\pagenumbering{gobble}
\title{Computation of the spectrum of $\text{dc}^2$-balanced codes
\thanks{Kees A. Schouhamer Immink is with Turing Machines Inc, Willemskade 15d, 3016 DK Rotterdam, The Netherlands. E-mail: immink@turing-machines.com. }
\thanks{Kui Cai is with Singapore University of Technology and Design (SUTD), 8 Somapah Rd, 487372, Singapore. E-mail: cai\_kui@sutd.edu.sg.}
\thanks{This work is supported by Singapore Agency of Science and Technology (A*Star) PSF research grant and SUTD-ZJU grant ZJURP1500102}
}
\author{Kees A. Schouhamer Immink and Kui Cai}
\maketitle

\begin{abstract} We apply  the central limit theorem for deriving approximations to the auto-correlation function and power density function (spectrum) of second-order spectral null ($\text{dc}^2$-balanced) codes. We show that the auto-correlation function of $\text{dc}^2$-balanced codes can be accurately approximated by a cubic function. We show that the difference between the approximate and exact spectrum is less than 0.04~dB for codeword length $n=256$. 
\end{abstract}
%
%
\section{Introduction}
Spectral null, or {\em dc-balanced}, codes have been applied in cable transmission~\cite{Cat},~\cite{Fr7}, magnetic recording~\cite{Ng0}, and optical recording systems~\cite{Ca9},~\cite{I24}. Spectral null codes have recently been advocated in visible light communications (VLC) systems, where light intensity of solid-state light sources, mostly LEDs, are varied~\cite{Ndj}. It is desirable that the intensity variation of the light is invisible to the users, that is, annoying {\em flicker} should be mitigated~\cite{Ohh}. This requirement implies that the spectrum of the modulated signal should not contain low-frequency components. Light sources are usually connected to the AC power grid, and therefore generate interference components at 50, 60~Hz, or the higher harmonics. Rejection of these interfering components can easily be accomplished by high-pass filtering, but in order not to degrade the wanted communication signal by this filtering, low-frequency components should be absent in the modulated signal. Three types of dc-balanced codes, the {\em Manchester code} (bi-phase), a {\em 4B6B code}, and a {\em 8B10B code}, have been adopted in VLC standard IEEE 802.15.7-2011~\cite{Raj} for flicker mitigation and dimming control~\cite{Ndj},~\cite{Wa4}.

Higher-order spectral null codes, such as $\text{dc}^2$-balanced codes, offer a greater rejection of the low-frequency components than regular dc-balanced codes~\cite{Im7}. Constructions of higher-order spectral null codes have been presented in for example~\cite{Ta8}, \cite{Xin}, \cite{Ya2}, \cite{Ska}, \cite{Xi1}, \cite{Rot}. Spectral properties of higher-order spectral null codes have been published for small values of the codeword length $n$~\cite{Im7}. For larger values of $n$, Immink and Cai~\cite{I77} have presented simple expressions for approximating the auto-correlation function and spectrum of higher-order spectral null codes. We apply statistical arguments for deriving improved approximations to the auto-correlation function and spectrum of $\text{dc}^2$-balanced codes. 

Section~\ref{spec} commences with background on $\text{dc}^2$-balanced codes. In Section~\ref{secautocor}, we derive an approximation to the auto-correlation function and spectrum of $\text{dc}^2$-balanced codes for asymptotically large values of the codeword length $n$ by counting $\text{dc}^2$-balanced codewords using the central limit theorem. Approximations for asymptotically large $n$ will be discussed in Section~\ref{secapprox}. In Section~\ref{secappraisal}, we appraise the spectral performance of $\text{dc}^2$-balanced codes. Section~\ref{secconc} shows our conclusions.
\section{Background on $\text{dc}^2$-balanced Block Codes}\label{spec}

Let the $n$-bit codeword $\bfx = (x_1, x_2,\ldots, x_n)$ over the binary symbol alphabet ${\cal Q }$ =$\{0, 1\}$, be a member of a codebook $S$. The encoder emits codewords from $S$ randomly and independently (i.i.d.). The auto-correlation function, $\rho(i)$, of a sequence of codeword symbols is given by~\cite{Bos},~\cite{Pie},~\cite{Jus}
\beq \rho(i) = \frac{1}{n|S|} \sum_{\bfx \in S} \sum_{j=1}^{n-i} x'_j x'_{j+i}, \,\, 0 \leq i \leq n-1 , \label{eqrhoi} \eeq
where $|S|$ denotes the cardinality of $S$ and $x'_i=2x_i-1$, $x'_i \in \{-1,1\}$, is the bipolar representation of $x_i$. If both $\bfx$ and its inverse $ \bar\bfx$ are members of $S$, then the power spectral density (psd), in short {\em spectrum}, versus frequency $\omega$ of the emitted symbol sequence is 
\beq H(\omega)  = 1 + 2 \sum_{i=1}^{n-1}\rho(i)\cos (i\omega) . \label{eqspec1a}\eeq
A regular `full-set' dc-balanced block code comprises all possible codewords that have equal numbers of 0's and 1's ($n$ even). Franklin and Pierce~\cite{Fr7} showed that the spectrum of a full-set dc-balanced block code has a null at the zero frequency, that is, $H(0) = 0$. $\text{Dc}^2$-balanced spectral null codes are dc-balanced codes that satisfy a second condition, namely
\beq H(0) = H(0)^{(2)}(0) =  0 \label{eqdercond}, \eeq
where $H^{(2)}(0)$ denotes the second derivative of $H(\omega)$ at $\omega=0$. Note that the above frequency domain conditions imply, see (\ref{eqspec1a}), that
\beq \sum_{i=1}^{n-1}\rho(i)=-\frac{1}{2} \mbox{{ \rm and }}\sum_{i=1}^{n-1}i^2\rho(i)=0.\label{eqchecks} \eeq
A codeword, $\bfx$, is $\text{dc}^2$-balanced if it satisfies~\cite{Im7},~\cite{Mon}
\beq \sum_{i=1}^n x_i=\frac{n}{2} \mbox{ and } \sum_{i=1}^n ix_i = \frac{n(n+1)}{4} . \label{eq_def}\eeq
A block code comprising a full set of $\text{dc}^2$-balanced codewords, denoted by $S_2$, is defined by
\beq S_2 = \left \{ \bfx \in {\cal Q}^n : \sum x_i=\frac{n}{2}; \sum ix_i = \frac{n(n+1)}{4} \right \} .  \eeq
The set $S_2$ is empty if $n \mod 4 \neq 0$~\cite{Im7}. Let $\bfx \in S_2$ then its reverse $\bfx_r= (x_n, \ldots, x_1) \in S_2$, since for a $\bfx\in S_2$
\beq \sum_{i=1}^n i x_i =  \sum_{i=1}^n (n+1-i) x_i = \frac{n(n+1)}{4} . \label{eqreverse} \eeq
A useful metric of the low-frequency spectral content, denoted by  $\chi$, called {\em Low Frequency Spectral Weight} (LFSW)~\cite{Xi1}, is the first non-zero coefficient of the Taylor expansion of (\ref{eqspec1a}), that is, 
\beq H (\omega) \sim \chi \omega^4, \,\, \omega \ll 1 \label{eqchi2} .\eeq 
We derive from (\ref{eqspec1a}) that
\beq \chi = \frac{1}{12} \sum_{i=1}^{n-1} i^4 \rho(i) . \label{eqchiK} \eeq 
The number of $\text{dc}^2$-balanced codewords, denoted by $N_{\text{dc}^2}=|S_2|$, for asymptotically large $n$, equals~\cite{Rot},~\cite{Pr1}
\beq N_{\text{dc}^2} \sim \frac{4\sqrt{3}} {\pi n^2} 2^n ,\,\,\, n \mod 4=0, \,\, n \gg 1. \label{eqNtext} \eeq
In the range $n<256$ we have found experimentally that a better approximation is found by applying a small correction term, namely
\beq N_{\text{dc}^2} \sim \frac{4\sqrt{3}} {\pi n^2} 2^n\left(1-\frac{1.211}{n}\right ),\,\,\, n \mod 4=0,\,\,n \gg 1.\label{eqNtext1}\eeq
We consider here the spectral properties of full-set block codes, that is, $S_2$ denotes the set of all possible words, $\bfx$, that satisfy condition (\ref{eq_def}). Finding an expression of the spectral properties of a full-set $S_2$ for large values of $n$ is an open problem as the computation requires the evaluation of (\ref{eqrhoi}) for each $\bfx \in S_2$~\cite{Xin}. In the next section, our main contribution, we address an alternative method, which is based on statistical analysis, which gives a simple and good approximation to the spectrum. 
\section{Auto-correlation function}\label{secautocor}
Let $\bfx$ be a codeword in $S_2$, and let $i_0$ and $i_1$, $i_0\neq i_1$, $ 1 \leq i_0,i_1 \leq n$, be two (different) index positions in the codeword $\bfx$. Then, we obtain for the average correlation, denoted by $r(i_0,i_1)$, between the symbols at positions $i_0$ and $i_1$ averaged over all codewords $\bfx \in S_2$, 
\begin{eqnarray}
r(i_0,i_1) &=& \frac{1}{N_{\text{dc}^2}}\sum_{\bfx \in S_2} (2x_{i_0}-1)(2x_{i_1}-1) \nonumber \\
&=& \frac{N_{\text{dc}^2}(x_{i_0}=x_{i_1}) - N_{\text{dc}^2} (x_{i_0}\neq x_{i_1}) } {N_{\text{dc}^2}}, 
\end{eqnarray}
where $N_{\text{dc}^2}(A)$ denotes the number of $\text{dc}^2$-balanced codewords $\bfx$ that satisfy condition $A$. Then, using (\ref{eqrhoi}) and (\ref{eqrhoi}), we find the auto-correlation function
\beq \rho(i) = \frac{1}{n}\sum_{j=1}^{n-i}  r(j,j+i)  , \,\, 0 \leq i \leq n-1 .\label{eqrhoix2}\eeq
For reasons of symmetry, we have
\begin{eqnarray} r(i_0,i_1) &=& \frac{2 N_{\text{dc}^2} (x_{i_0}=x_{i_1})}{N_{\text{dc}^2}} - 1 \nonumber\\
&=& \frac{4N_{\text{dc}^2} (x_{i_0}=x_{i_1}=1)}{N_{\text{dc}^2}} - 1  . \label{eqri0i1}\end{eqnarray}
By using the central limit theorem, we compute below an approximation to the number of $\text{dc}^2$-balanced codewords that have a `1' at positions $i_0$ and $i_1$, $N_{\text{dc}^2}(x_{i_0}=x_{i_1}=1)$, for asymptotically large values of $n$. 
\subsection{Counting of codewords using the central limit theorem}
The number of $\text{dc}^2$-balanced codewords, $\bfx$, $N_{\text{dc}^2}(x_{i_0}=x_{i_1}=1)$, that is required for computing the auto-correlation function using (\ref{eqrhoix2}) and (\ref{eqri0i1}), can be computed using generating functions. For very large $n$, however, this rapidly becomes an impractically cumbersome exercise, and an efficient alternative method is considered a desideratum. 

To that end, we exploit the central limit theorem by regarding the integer variables $x_i \in \{0,1\}$ as i.i.d. binary random variables whose numerical outcomes `0' or `1' are equally likely. We define the stochastic variables $c$ and $p$ by
\beq  c = x_1 + x_2 + \cdots + x_n  \label{eqs1}\eeq
and
\beq  p = x_1 + 2x_2 + \cdots + nx_n,  \label{eqp1}\eeq
where $x_{i_0}=x_{i_1}=1$, $i_0, i_1 \in \{1,\ldots, n\}$. 

The central limit theorem~\cite{Fla}, Chapter~8, states that for asymptotically large $n$ the distribution of the stochastic variables $c$ and $p$, which are obtained by summing a large number, $n$, of independent stochastic variables, approaches a two-dimensional Gaussian distribution. 

Let $E[.]$ denote the expected value operator for all possible codewords in $S_2$. Let the parameters $\mu_c=E[c]$ and $\mu_p=E[p]$ denote the average of $c$ and $p$, and let $\sigma_c^2=E[(c-\mu_c)^2]$ and $\sigma_p^2=E[(p-\mu_p)^2]$ denote the variance of $c$ and $p$. The parameter $r$ denotes the linear correlation coefficient between the random variables $c$ and $p$. Then the bi-variate Gaussian distribution, denoted by $G(c,p)$, is given by
\beq  G(c,p) = \frac {1} {2\pi\phi_1} e^{-\phi(c,p)}, \eeq
where 
\beq \phi^2_1 = \sigma^2_c\sigma^2_p (1-r^2),  \label{eqfcp1}\eeq
\beq \phi(c,p) = \frac{1}{2(1-r^2)}f(c,p) , \label{eqfcp2}\eeq
and
$$ f(c,p) = \left (\frac{c-\mu_c}{\sigma_c} \right)^2 + \left(\frac{p-\mu_p}{\sigma_p}\right)^2 -\frac{2r(c-\mu_c)(p-\mu_p)} {\sigma_c\sigma_p} .$$ 
We have 
\beq E[x_i]=E[x^2_i]= \frac{1}{2}, \mbox{ and } E[x_ix_j]=\frac{1}{4}, \,\, i \neq i_0, i_1 , \label{eqexps} \eeq
\beq E[x_i]=E[x^2_i]=1, \, i=i_0,i_1,\, E[x_ix_j]=\frac{1}{2},\, i,j=i_0,i_1 , \nonumber  \eeq
and $E[x_{i_0}x_{i_1}]=1$. We may find after a routine computation using (\ref{eqexps})  that  
$$
\mu_c = E\left[\sum_{i=1}^n x_i \right] = \sum_{i=1}^n E[x_i] = \frac{n-2}{2} +2 $$ 
and similarly
$$\mu_p=E\left [\sum_{i=1}^n ix_i \right]=\sum_{i=1}^n  E\left [ix_i \right]= \frac{n(n+1)}{4}+\frac{i_0+i_1}{2}.$$ 
The variances $\sigma_c^2 $, $\sigma_p^2$, and the correlation coefficient $r$ can be found without too much difficulty:
\beq \sigma^2_c = E \left[\sum_{i=1}^n (x_i-\mu_c)^2 \right]= \frac{n-2}{4}, \eeq
\begin{eqnarray} \sigma_p^2 &=& E \left[\sum_{i=1}^n (ix_i-\mu_p)^2 \right] \nonumber\\
&=&  \frac{n(n+1)(2n+1)}{24} - \frac{i_0^2+i_1^2}{4}, \end{eqnarray}
and
\begin{eqnarray}
r^2 &=& \frac{E[\sum_{i=1}^n (x_i-\mu_c). \sum_{i=1}^n (ix_i-\mu_p) ]}{\sigma^2_c \sigma_p^2}  \nonumber\\
&=& \frac{3}{2(n-2)}\frac{(n^2+n-2(i_0+i_1))^2}{2n^3+3n^2+n-6(i_0^2+i_1^2)} . \label{eqrho2}
\end{eqnarray}
The total number of $n$-sequences with $x_{i_0}=x_{i_1}=1$, equals $2^{n-2}$, so that for asymptotically large $n$, the number of $n$-sequences versus $c$ and $p$, denoted by $N(c,p;x_{i_0}=x_{i_1}=1)$, can be approximated by 
\beq N(c,p;x_{i_0}=x_{i_1}=1) \sim 2^{n-2} G(c,p) = \frac {2^n} {8\pi\phi_1} e^{-\phi(c,p)} . \label{eqfcp0}\eeq
A $\text{dc}^2$-balanced codeword satisfies, by definition, the conditions, see (\ref{eq_def}), $c=n/2$ and $p=n(n+1)/4$. Then, $N_{\text{dc}^2}(x_{i_0}=x_{i_1}=1)$ is found after substituting $c=n/2$ and $p=n(n+1)/4$ into (\ref{eqfcp0}). We find
\beq N_{\text{dc}^2}(x_{i_0}=x_{i_1}=1) \sim \frac {2^{n}} {8\pi\phi_1} e^{-\phi_2}, \label{eqvarphi}\eeq
where, see (\ref{eqfcp1}) and (\ref{eqfcp2}),
\begin{eqnarray} 
\phi_2 &=& \phi \left(c=\frac{n}{2},p=\frac{n(n+1)}{4} \right) \nonumber\\ 
&=&\frac{1}{2(1-r^2)} \left \{\left(\frac{1}{\sigma_c}\right)^2 + \left(\frac{i_0+i_1}{2\sigma_p}\right)^2 -\frac{ r(i_0+i_1)}{\sigma_c\sigma_p} \right \}\nonumber\\ 
&=& \frac{4\sigma^2_p +\sigma^2_c(i_0+i_1)^2-4r\sigma_c\sigma_p(i_0+i_1) }{8\phi_1}.\label{eqfcp} \end{eqnarray}
After combining (\ref{eqNtext}), (\ref{eqri0i1}), and (\ref{eqvarphi}), we obtain
\beq r(i_0,i_1)  = \frac {n^2} {\sqrt{192}\, \phi_1} e^{-\phi_2} - 1 . \eeq 
In order to reduce the clerical work and offer more insight, we define the four (real) variables
$$ \gamma=12n[(i_0-n-1)i_0 + (i_1-n-1)i_1],$$ $$ \delta=(i_0-i_1)^2,$$ $$ r_1=-\frac{1}{n^4}(8 n^3+13n^2+4n+\gamma-12\delta),$$
and
$$ r_2=\frac{1}{8n^3}[12{n}^2+4n+\gamma-6(n+2)\delta]. $$
With some effort we find the expressions
\beq \phi_1^2= \frac{n^4}{192}(1+r_1) \eeq 
and
\beq \phi_2 = \frac{8}{n} \frac{1+r_2}{1+r_1} . \eeq
We finally obtain
\begin{eqnarray}
r(i_0, i_1)  &=& \frac {n^2} {\sqrt{192}\, \phi_1} e^{-\phi_2} -1 \nonumber\\
&=& \frac{1}{\sqrt{1+r_1}}e^{-\frac{8}{n} \frac{1+r_2}{1+r_1}} -1 , \label{eqri0i1a}\end{eqnarray}
where we can easily verify, since $\bfx$ and $\bfx_r \in S_2$, see (\ref{eqreverse}), that
\beq r(i_0,i_1) = r(n+1-i_0,n+1-i_1) .\eeq 
The auto-correlation function, $\rho(i)$, is found using (\ref{eqrhoix2}). In the next subsection, we show results of computations. 
\subsection{Results of computations}
\begin{figure} \centerline{\psfx{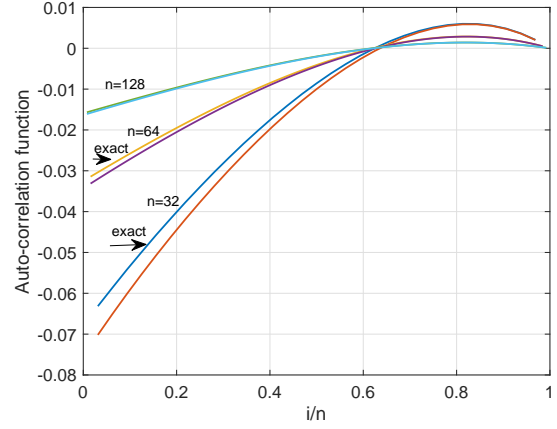}{8cm}} \caption{\protect\small Auto-correlation functions, $\rho(i)$ (estimate, using (\ref{eqrhoix2}) and (\ref{eqri0i1a})) and $\hat \rho(i)$ (exact, full set), versus $i/n$ for $n=32, 64$ and 128.} \label{figspec2} \end{figure}

By invoking (\ref{eqrhoix2}) and (\ref{eqri0i1a}) we are now able to compute an estimate of the auto-correlation function $\rho(i)$. Figure~\ref{figspec2} shows results of computations for $n=32$, 64, and~128. As a comparison we plotted the exact auto-correlation function of a full set of $\text{dc}^2$-balanced sequences, denoted by $\hat\rho(i)$, which was computed using an enumeration technique and generating functions~\cite{Im7}. 
\begin{figure} \centerline{\psfx{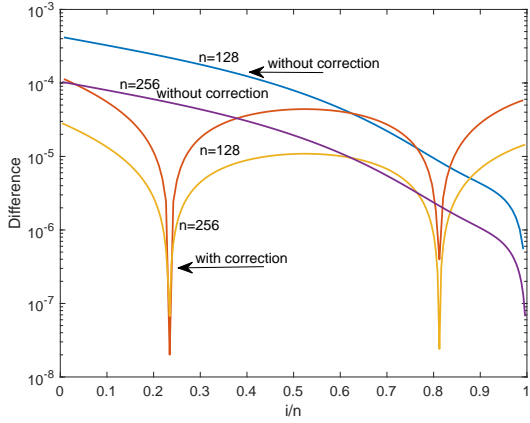}{8cm}} \caption{\protect\small Difference, with and without correction, between estimated and exact auto-correlation functions, $|\hat\rho(i)-\rho(i)|$, versus $i/n$ for $n=128$ and 256.}\label{figrhodiff} \end{figure}

The accuracy of the approximate auto-correlation function, $\rho(i)$, cannot easily be determined from Figure~\ref{figspec2} for the larger values of $n$. Figure~\ref{figrhodiff}, Curve `without correction', shows $|\hat\rho(i)-\rho(i)|$, the difference between the two auto-correlation functions versus $i/n$ for the selected $n=128$ and~256. We notice that the difference between the two functions decreases with increasing $i$ and $n$. For $n=256$ in the whole range the difference, $|\hat\rho(i)-\rho(i)|$, is less than $10^{-4}$.

Although the difference, $|\hat\rho(i)-\rho(i)|$, is relatively small, especially for larger $n$, see Figure~\ref{figrhodiff}, the `checks', see (\ref{eqchecks}), $\sum \rho(i)=-1/2$ and $\sum i^2\rho(i)=0$, which accumulate the small error differences, are not necessarily satisfied. We have observed that with increasing $n$ that $\sum \rho(i)+1/2$ is converging to zero (as it should), while $\sum i^2\rho(i)$ is not. As a result, the spectra, computed using $\rho(i)$ do not satisfy the spectral conditions (\ref{eqdercond}). 

We propose to add a small correction term to $\rho(i)$ so that both `checks', $\sum \rho(i)=-\frac{1}{2}$ and $\sum i^2\rho(i)=0$, are satisfied. We add to $\rho(i)$ the correction term $a+bi$, where the (real) parameters, $a$ and $b$, are chosen such that $\sum_i (\rho(i)+a+bi)=-\frac{1}{2}$ and $\sum_i i^2(\rho(i)+a+bi)=0$. Define $$a_0=\sum \rho(i)+\frac{1}{2}$$ and $$a_1=\sum i^2\rho(i),$$ then we find two linear equations with two unknowns, $a$ and $b$, namely
$$ \sum_{i=1}^{n-1} \left (\rho(i)+a+bi \right) = a_0-\frac{1}{2} + na + b\sum_{i=1}^{n-1} i= -\frac{1}{2}   $$
and
$$ \sum_{i=1}^{n-1} i^2 \left (\rho(i)+a+bi \right) = a_1 + a\sum_{i=1}^{n-1} i^2 + b\sum_{i=1}^{n-1} i^3 =0.  $$
After solving the above system, where we substitute the well-known expressions for $\sum i^k$, $k=1,2,3$, we obtain 
\beq a =-3 \frac{n(n-1)a_0-2a_1}{n(n-1)(n-2)} \label{eqaa1}\eeq 
and 
\beq b = 2\frac{n(2n-1)a_0-6a_1}{n^2(n-1)(n-2)}. \label{eqaa2}\eeq
For example, for $n=128$, we find that $a_0=-0.0156$ and $a_1=-22.21$. So that $a =0.0003063$ and $b=-0.0000029$. The result of the correction can be seen in Figure~\ref{figrhodiff}, curves `with correction', for $n=128$ and 256. We notice in the range $i/n<0.6$ a significant improvement in the accuracy of the estimate of the auto-correlation function. 
\subsection{Further approximations for asymptotically large $n$}\label{secapprox}
\begin{figure} \centerline{\psfx{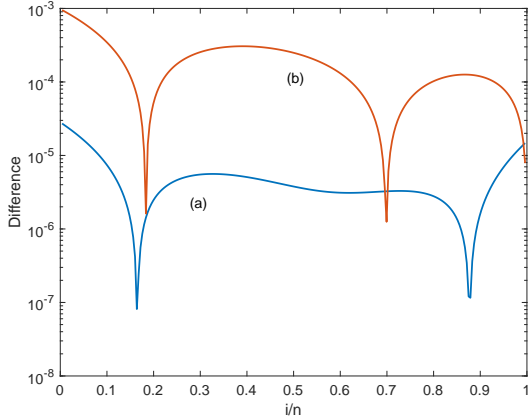}{8cm}} \caption{\protect\small Deviation between exact and estimated auto-correlation function a) $|\rho'(i)-\hat\rho(i)|$ and b) $|\rho_a(i)-\hat\rho(i)|$ versus $i/n$ for $n=256$.} \label{figdiff2} \end{figure}

\begin{figure} \centerline{\psfx{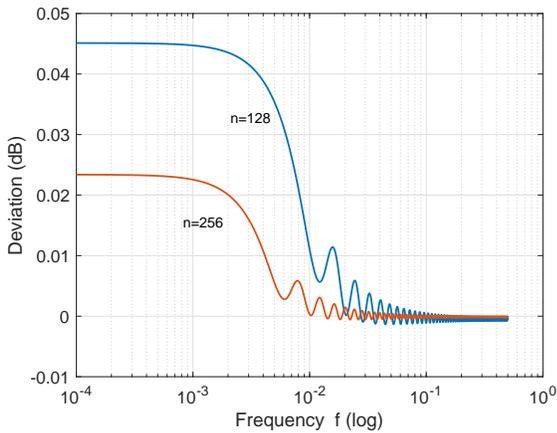}{8cm}} \caption{\protect\small Deviation between exact and estimated spectrum $H'(\omega)/\hat H(\omega)$ (dB) versus $\omega$ for $n=128$ and 256.} \label{figspectradc2} \end{figure}

With (\ref{eqri0i1a}) we can straightforwardly compute the auto-correlation function $\rho(i)$ and spectrum $H(\omega)$. In this section, we attempt to approximate $\rho(i)$ for asymptotically large $n$, which might offer more insight in the trade-offs between redundancy and spectral properties. We apply to (\ref{eqri0i1a}) the well-known series approximations
$$ \frac{1}{\sqrt{1+x}} = 1-\frac{x}{2} + \frac{3}{8}x^2- \frac{5}{16}x^3 + \cdots $$
and
$$ \frac{1}{1+x} = 1-x+x^2-x^3 + \cdots.$$ 
We have experimented with the various options available for trading accuracy versus simplicity of the expression, and propose  
\beq r(i_0, i_1) \sim -\frac{8}{n}(1+r_2)-\frac{r_1}{2}, \,\, n \gg1. \eeq
Using (\ref{eqrhoix2}), we obtain 
\beq \rho(i) \sim \frac{n-i}{2n^5}\left(12i^2+4ni^2+4n^2i-4n^3+n^2+4n \right) . \eeq
Then, after deleting the smallest terms, we obtain the simple cubic function
\beq \rho(i) \sim \frac{2}{n^4} (n-i)(i^2+in-n^2), \eeq
which can be rewritten as
\beq \rho(i) \sim \frac{2}{n^4}(n-i)(i-c_0n)(i-c_1n),\eeq
where $c_{0,1}=(-1 \mp \sqrt{5})/2$. The checks (\ref{eqchecks}) for the above $\rho(i)$ yield
$$a_0=\frac{1}{2}+\sum\rho(i)=\frac{1}{n}-\frac{1}{2n^2}$$ and $$a_1=\sum i^2\rho(i)=-\frac{1}{6}+\frac{1}{6n^2}.$$  
In order to satisfy both checks (\ref{eqchecks}), we add to $\rho(i)$ the correction term $a+bi$, and define 
\beq \rho'(i) = \frac{2}{n^4} (n-i)(i-c_0n)(i-c_1n) +a+bi, \label{eqfin}  \eeq
where after using (\ref{eqaa1}) and (\ref{eqaa2}), we obtain
$$ a = -\frac{6n^2-n+2}{2(n-2)n^3} \sim - \frac{3}{n^2} $$ and $$ b= \frac{4n^3-2n^2+n-2}{n^4(n-1)(n-2)} \sim \frac{4}{n^3}.  $$
Note that $a$ and $b$ are relatively small terms in (\ref{eqfin}) for asymptotically large $n$.
Figure~\ref{figdiff2} shows the difference between exact and estimated auto-correlation function $|\rho'(i)-\hat\rho(i)|$ versus $i/n$ for $n=256$.
As a final proof of the pudding, we compare the (exact) spectrum of full set codewords versus the spectrum, denoted by $H'(\omega)$, which is computed using the above approximated auto-correlation function $\rho'(i)$. The difference, $H'(\omega)/\hat H(\omega)$ (dB), between the spectrum, $H'(\omega)$, computed using $\rho'(i)$, and the exact spectrum, $\hat H(\omega)$, of full set codewords, which was computed using generating functions, is plotted in Figure~\ref{figspectradc2}. We may observe that the difference between the two spectra is very small, less than 0.05~dB for $n=128$ and less than 0.03~dB for $n=256$. 

\begin{figure} \centerline{\psfx{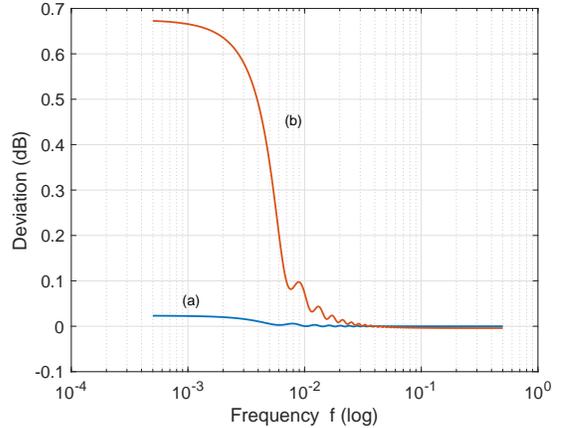}{8cm}} \caption{\protect\small Deviation between exact and estimated spectrum a) $H'(\omega)/\hat H(\omega)$ (dB) and b) $H_a(\omega)/\hat H(\omega)$ (dB) versus $\omega$ for $n=256$.} \label{figspecdiff2} \end{figure}
\begin{table}\caption{LFSW $\chi'$ and $\hat \chi$ versus $n$.} 
$$\begin{array}{r|r|r} \hline n & \chi' & \hat \chi\\ \hline 32 & 1629.48  & 1576.72 \\64 & 24723.13  & 24250.79 \\
128 & 384339.75  & 380367.61 \\256 & 6057889.79  & 6025352.62 \\\hline \end{array}$$ \label{tabchis}\end{table} 
The LFSW metric, $\chi'$ is, using (\ref{eqchiK}),
\beq \chi' = \frac{1}{12} \sum_{i=1}^{n-1} i^4 \rho'(i) \sim \frac{n^4}{720}\left (1+ \frac{4}{n} \right)  . \eeq 
Table~\ref{tabchis} shows $\chi'$ for selected values of $n$, where as a comparison we have listed the LFSW of full set $\text{dc}^2$-balanced codes, denoted by $\hat\chi$. We may notice that for $n=256$ the difference between $\chi$ and $\hat \chi$  is less than half a percent. 
\subsection{Comparison with prior art}
In \cite{I77}, it is postulated that the auto-correlation of $\text{dc}^2$-balanced spectral null codes, denoted by $\rho_a(i)$, can be modelled by the simple parabola's equation
\beq \rho_a(i) =\beta(i+\alpha)(i-n) \label{eqK2},   \eeq 
where the (real) parameters $\alpha$ and $\beta$ are given by
$$ \alpha = -\frac{3n^2-2}{5n}$$ and $$ \beta={\frac {-15}{(n-1)(n-2)(4n+3)}}.$$ 
It has been shown in~\cite{I77} that the parabola's equation (\ref{eqK2}) is an accurate approximation to the exact correlation function of full-set $\text{dc}^2$-balanced spectral null codes. Figure~\ref{figdiff2}, Curve (b), shows the difference between exact and estimated auto-correlation function $|\rho_a(i)-\hat\rho(i)|$ versus $i/n$ for $n=256$. We notice that the newly developed $\rho'(i)$, Curve (a), is almost an order more accurate than (\ref{eqK2}) presented in the prior art. Figure~\ref{figspecdiff2}, Curve~b, shows that the quotient of the exact spectrum and the one based on prior art (\ref{eqK2}), $\hat H(\omega)/H_a(\omega)$, is for $n=256$ less than 0.7~dB, and also here we notice that the newly developed theory is more than an order more accurate.

\section{Appraisal of spectral performance}\label{secappraisal}
A system designer is usually confronted with a restricted redundancy budget, so that with a given redundancy the designer searches for a balanced code that offers the best rejection of low-frequency components. In this section, we compare the spectral performance of regular dc-balanced codes with that of $\text{dc}^2$-balanced codes. We start with a summary of properties of dc-balanced codes.
\subsection{Codes with a first-order spectral null}
Let the codeword length of a regular full-set dc-balanced code be denoted by $n_1$, $n_1$ even. Each codeword has an equal number of 0's and 1's, so that the number of available dc-balanced codewords, denoted by $N_{\text{dc}}$, is simply~\cite{I57} 
\beq N_{\text{dc}} = {{n_1} \choose {n_1/2}} \sim \frac{1}{\sqrt{\frac{\pi}{2} n_1}} 2^{n_1}, \,\, n_1 \gg 1.  \label{eqNdc}\eeq 
The auto-correlation function, $\rho_1(i)$, and the spectrum, $H_1(\omega)$, of dc-balanced codes is~\cite{Fr7}  
\beq \rho_1(i) =  \frac{1}{n_1(n_1-1)} (i-n_1)\eeq
and
\beq H_1(\omega)={{n_1} \over {n_1-1}}\left \{1-\left({{{\sin \frac{n_1 \omega}{2}}}\over {{n_1\sin\frac{\omega}{2}}}}\right )^2\right\} . \label{eqspe80} \eeq
At the very low-frequency end, we have\cite{Xi1} 
\beq H_1(\omega) \sim \chi_1 \omega^2, \,\, \omega \ll 1, \eeq
where
\beq \chi_1 = \frac{n_1(n_1+1)}{12}  .\eeq
\subsection{Performance comparison}
We compare the spectral content of dc-balanced versus that of $\text{dc}^2$-balanced codes, where we assume that both types of codes have the same redundancy. Let $R$ and $R_1$ denote the maximum information rate of a $\text{dc}^2$-balanced code or $\text{dc}$-balanced of length $n$ and $n_1$, respectively, then we have, using (\ref{eqNtext}),  
\begin{eqnarray}
R &=& \frac{1}{n} \log_2 N_{\text{dc}^2} = \frac{1}{n} \log_2 \frac{4\sqrt{3}} {\pi n^2} 2^n\nonumber\\
&=& 1-\frac{1}{n} \log_2 \frac{\pi n^2}{4\sqrt{3}}  \end{eqnarray}
and, using (\ref{eqNdc}),
\beq R_1 = 1-\frac{1}{2n_1} \log_2 \frac{\pi}{2} n_1 .  \eeq
\begin{table}\caption{Code length $n$ and $n_1$ for $R=R_1$.} $$\begin{array}{r|r|r} \hline R=R_1 & n_1 & n\\ \hline 
0.90  &     28 &     132\\0.92  &     38 &     172\\0.94  &     54 &     248\\0.96  &     90 &     408\\
0.98  &    210 &     932\\\hline \end{array}$$ \label{tabn1n}\end{table} 
Table~\ref{tabn1n} shows a few examples of the codewords length $n$ and $n_1$ for which dc-balanced and $\text{dc}^2$-balanced codes have equal redundancy, respectively, that is, $R=R_1$. In the range shown in Table~\ref{tabn1n}, the codeword length $n$ of a $\text{dc}^2$-balanced code is approximately a factor of 4.5 larger than the codeword length $n_1$ of a dc-balanced code for achieving the same rate $R=R_1$.
\begin{figure} \centerline{\psfx{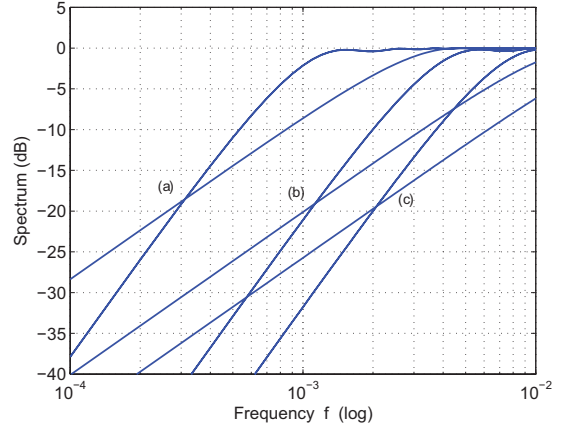}{8cm}} \caption{\protect\small Spectra of dc-balanced and $\text{dc}^2$-balanced codes  with the same redundancy versus frequency for a) $R=R_1=0.98$, b) $R=R_1=0.94$, and c) $R=R_1=0.90$, see also Table~\ref{tabn1n}. The points of intersection are around -20~dB. \label{figspec3}}\end{figure}
Figure~\ref{figspec3} shows three examples of spectrum pairs of dc-balanced and $\text{dc}^2$-balanced codes with the same redundancy versus frequency for a) $R=R_1=0.98$, b) $R=R_1=0.94$, and c) $R=R_1=0.90$, see also Table~\ref{tabn1n}. We may notice the points of intersection of the spectra of dc-balanced and $\text{dc}^2$-balanced codes. A further perusal of the diagram reveals that the points of intersection are at around -20~dB, which implies that $\text{dc}^2$-balanced codes are to be preferred when a low-frequency spectral suppression is required better than around -20~dB. Additional computations show that this `20~dB rule' applies to all codes with a rate larger than 0.75. 
\section{Conclusions}\label{secconc}
By applying the central limit theorem, we have derived an approximate expression for the auto-correlation function and spectrum of full-set $\text{dc}^2$-balanced codes for asymptotically large values of the codeword length $n$. We have shown that the auto-correlation function of $\text{dc}^2$-balanced codes can be accurately approximated by a simple cubic function. We have compared the approximate spectrum with the exact spectrum of full set $\text{dc}^2$-balanced codes. We have shown that the difference between the approximated and exact spectrum is less than 0.04~dB for $n=256$. 


\begin{thebibliography}{99}

\bibitem{Cat} 
K. W. Cattermole, ``Principles of Digital Line Coding,'' {\em Int. Journal of Electronics,\/} vol. 55, pp. 3-33, July 1983.
\bibitem{Fr7} 
J. N. Franklin and J. R. Pierce, ``Spectra and Efficiency of Binary Codes without DC,'' {\em IEEE Trans. Commun.,\/} vol. COM-20, pp. 1182-1184, Dec. 1972.
\bibitem{Ng0} 
Y. Ng, K. Cai, K. S. Chan, M. R. Elidrissi, M. Y. Lin, Z. Yuan, C .L. Ong, and S. Ang, ``Signal Processing for Dedicated Servo Recording System,'' {\em IEEE Trans. Magn.,\/} vol. 51, no. 10, Oct. 2015. 
\bibitem{Ca9} 
K. Cai, K. A. S. Immink, M. Zhang, and R. Zhao, ``Design of Spectrum Shaping Codes for High-Density Data Storage,'' {\em Trans. on Consumer Electronics\/}, vol. CE-63, pp. 477-482, Nov. 2017. 
\bibitem{I24} 
K. A. S. Immink, ``Spectral Null Codes,'' {\em IEEE Trans. Magn.,\/} vol. MAG-26, no. 2, pp. 1130-1135, March 1990.
\bibitem{Ndj} 
A. R. Ndjiongue, H. C. Ferreira, and T. M. N. Ngatched, {\em Visible Light Communications (VLC) Technology\/}, 
Wiley Encyclopedia of Electrical and Electronics Engineering, 2015.
\bibitem{Ohh} 
M. Oh, ``A Flicker Mitigation Modulation Scheme for Visible Light Communications'',  2013 15th International Conference on Advanced Communications Technology (ICACT), PyeongChang, South Korea, Jan. 2013.
\bibitem{Raj} 
S. Rajagopal, R. D. Roberts, and S-K. Lim,
``IEEE 802.15.7 Visible Light Communication: Modulation Schemes and Dimming Support,'' {\em IEEE Communications Magazine,\/} vol. 50, No.3, pp. 72-82, March 2012. \bibitem{Wa4} 
Z. Wang, Q. Wang, W. Huang, and Z. Xu, {\em Visible Light Communications: Modulation and Signal Processing,\/} Wiley-IEEE Press, Jan 2018.
\bibitem{Im7} 
K. A. S. Immink and G. F. M. Beenker, ``Binary Transmission Codes with Higher Order Spectral Zeros at Zero Frequency,'' {\em IEEE Trans. Inform. Theory,\/} vol. IT-33, no. 3, pp. 452-454, May 1987.
\bibitem{Ta8} 
L. G. Tallini and B. Bose, ``On Efficient High-Order Spectral-Null Codes,'' {\em IEEE Trans. Inform. Theory,\/} vol. IT-45, no. 7, pp. 2594-2601, Nov. 1999.
\bibitem{Xin} 
Y. Xin and I. J. Fair, ``Algorithms to Enumerate Codewords for $\text{DC}^2$-constrained Channels,'' {\em IEEE Trans. Inform. Theory,\/} vol. IT-47, no. 7, pp. 3020-3025, Nov. 2001.
\bibitem{Ya2} 
 C. N. Yang, ``Efficient Encoding Algorithm for Second-Order Spectral-Null Codes Using Cyclic Bit Shift,'' {\em IEEE Transactions on Computers\/}, vol. 57, no. 7, pp. 876-888, July 2008.
\bibitem{Ska} 
V. Skachek, T. Etzion, and R. M. Roth, ``Efficient Encoding Algorithm for Third-order Spectral-Null Codes,'' {\em IEEE Trans. Inform. Theory,\/} vol. IT-44, pp. 846-851, March 1998.
\bibitem{Xi1} 
Y. Xin and I. J. Fair, ``A Performance Metric for Codes with a High-Order Spectral Null at Zero Frequency,'' {\em IEEE Trans. Inform. Theory,\/} vol. IT-50, no. 2, pp. 385-394, Feb. 2004.
\bibitem{Rot} 
R. M. Roth, P. H. Siegel, and A. Vardy, ``Higher-Order Spectral-Null Codes: Constructions and Bounds,'' {\em IEEE Trans. Inform. Theory,\/} vol. IT-40, pp. 1826-1840, Nov. 1994.
\bibitem{I77} 
K. A. S. Immink an K. Cai, ``Estimated Spectra of Higher-Order Spectral Null Codes,'' {\em IEEE Commun. Letters\/}, Oct. 2018.
\bibitem{Bos} 
B.S. Bosik, ``The Spectral Density of a Coded Digital Signal,'' {\em Bell Syst. Tech. J.,\/} vol. 51, pp. 921-932, April 1972.
\bibitem{Pie} 

G. L. Pierobon, ``Codes for Zero Spectral Density at Zero Frequency,'' {\em IEEE Trans. Inform. Theory,\/} vol. IT-30, no. 2, pp. 435-439, March 1984.
\bibitem{Jus} 

J. Justesen, ``Information Rates and Power Spectra of Digital Codes,'' {\em IEEE Trans. Inform. Theory,\/} vol. IT-28, no. 3, pp. 457-472, May 1982.
\bibitem{Mon} 

C. M. Monti and G. L. Pierobon, ``Codes with a Multiple Spectral Null at Zero Frequency,'' {\em IEEE Trans. Inform. Theory,\/} vol. IT-35, no. 2, pp. 463-472, March 1989.
\bibitem{Pr1} 

H. Prodinger, ``On the Number of Partitions of \{$1, \ldots ,n$\} into Two Sets of Equal Cardinalities and Equal Sums,'' 
{\em Canad. Math. Bull.,\/} vol. 25, no. 2, pp. 238-241, 1982.
\bibitem{Fla} 

P. Flajolet and R. Sedgewick, {\em Analytic Combinatorics}, ISBN 978-0-521-89806-5, Cambridge University Press, 2009. 
\bibitem{I57} 
 K. A. S. Immink and J. H. Weber, ``Very Efficient Balanced Codes,'' {\em IEEE Journal on Selected Areas of Communications\/}, vol. 28, pp. 188-192, 2010.
\end{thebibliography}
\end{document}